\documentclass[iop]{emulateapj}

\usepackage{graphicx}
\usepackage{lineno}
\usepackage{amssymb,amsmath}
\usepackage{color}
\usepackage{tablefootnote}

\shorttitle{Blazar LoS Voidiness}
\shortauthors{Furniss, Sutter, Primack,  \& Dom\'{i}nguez}

\begin{document}

\title{A Correlation between Hard Gamma-ray Sources and Cosmic Voids Along the Line of Sight}

\author{A.~Furniss\altaffilmark{1}, P.~M.~Sutter\altaffilmark{2,3,4}, J.~R.~Primack\altaffilmark{5}, and 
A.~Dom\'{i}nguez\altaffilmark{6}}
%\affil{Astronomy Department, University of California, Berkeley, CA 94720}

\altaffiltext{1}{Stanford University, Stanford, CA 94305, USA; {\color{blue} amy.furniss@gmail.com}}
\altaffiltext{2}{Sorbonne Universit\'{e}s, UPMC Univ Paris 06, UMR7095, Institut d'Astrophysique de Paris, F-75014, Paris, France}
\altaffiltext{3}{CNRS, UMR7095, Institut d'Astrophysique de Paris, F-75014, Paris, France}
\altaffiltext{4}{Center for Cosmology and AstroParticle Physics, Ohio State University, Columbus, OH 43210 USA}
\altaffiltext{5}{Physics Department, University of California, Santa Cruz, CA 95064 USA}
\altaffiltext{6}{Department of Physics and Astronomy, University of California, Riverside, CA 92521 USA}

\begin{abstract}
We estimate the galaxy density along lines of sight to hard extragalactic gamma-ray sources by correlating source positions on the sky with a void catalog based on the Sloan Digital Sky Survey (SDSS).  
Extragalactic gamma-ray sources that are detected at very high energy (VHE; $E>100$ GeV) or have been highlighted as VHE-emitting candidates in the $Fermi$ Large Area Telescope hard source catalog (together referred to as ``VHE-like" sources) are distributed along underdense lines of sight at the 2.4$\sigma$ level.  There is also a less suggestive correlation for the $Fermi$ hard source population (1.7$\sigma$).  A correlation between 10-500 GeV flux and underdense fraction along the line of sight for VHE-like and $Fermi$ hard sources is found at $2.4\sigma$ and $2.6\sigma$, calculated from the Pearson correlation coefficients of $r=0.57$ and $0.47$, respectively.  The preference for underdense sight lines is not displayed by gamma-ray emitting galaxies within the second $Fermi$ catalog, containing sources detected above 100 MeV, or the SDSS DR7 quasar catalog.  We investigate whether this marginal correlation might be a result of lower extragalactic background light (EBL) photon density within the underdense regions and find that, even in the most extreme case of a entirely underdense sight line, the EBL photon density is only 2\% less than the nominal EBL density.  Translating this into gamma-ray attenuation along the line of sight for a highly attenuated source with opacity $\tau(E,z) \sim 5$, we estimate that the attentuation of gamma-rays decreases no more than 10\%.  This decrease, although non-neglible, is unable to account for the apparent hard source correlation with underdense lines of sight.  %If the EBL density is indeed uniform, then this result would support a negligible to vanishing magnitude of the intergalactic magnetic field within voids, although more detailed calculations of the line-of-sight density and pair cascade interactions will be necessary to confirm this result.
\end{abstract}


\section{Introduction} 
%Blazars are active galaxies that have a jet aligned along with, or close to,
%the Earth line of sight \citep{urry}.   These sources emit broadband emission
%from radio to very high energy (VHE; $E\ge$100 GeV) gamma rays.  
%As a result of
%the highly peaked interaction cross-section and specific energy threshold of
%photon-photon pair production, 
The flux from extragalactic very high energy (VHE; $E\ge$100 GeV) gamma-ray sources is
absorbed by extragalactic background light (EBL) photons via pair production in
an energy- and distance-dependent manner \citep{nikishov,gould}.  % due to the highly 
%peaked interaction cross-section and specific energy threshold of photon-photon pair production.  
As a consequence of this absorption, the EBL produces an opacity $\tau(E,z)$ for VHE photons of 
observed energy $E$ from a source located at redshift $z$.  This opacity leads to an exponential flux attenuation $e^{-\tau(E,z)}$.

The EBL consists of all the evolving accumulated and reprocessed light from the UV to the far-IR and 
%JP: The EBL includes light from AGN and any other sources (such as hypothetical particle decay)
is difficult to measure directly due to strong foreground sources
\citep[e.g.,][]{hauser}.    A selection of the currently available models of the EBL
photon density include a semi-analytical model \citep{gilmore2012}, a model
based on observations of galaxy spectral energy distribution fractions \citep{dominguez}, and a
fitting, interpolating and multiwavelength backward evolution of cosmological
survey data \citep{franceschini}.  

These EBL models estimate photon densities at or above the conservative lower limits
derived from galaxy counts, and are each below the indirectly
set upper limits based on the detection of extragalactic VHE photons from
sources at z$\sim$0.1-0.2 (e.g., \citealt{aharonian2006,meyer2012}).   More recent use of
extragalactic gamma-ray sources for indirect studies of the EBL photon density
have confirmed an EBL density not much above the observational $z\sim0$ lower limits 
and confirmed the ``cosmic gamma ray horizon," defined as $\tau(E,z)=1$ 
\citep[e.g.,][]{fermiEBL,HESSEBL, dominguez2013}.  However, even with a low EBL photon density, a
number of distant blazars have recently been detected at energies that probe opacities between $\tau=2-5$  \citep{archambault,furniss1424, KUV, 3C279}.  
% despite the expected ``gamma-ray horizon" above $\tau=1$
%This gamma-ray horizon results from the absorption of the intrinsic VHE photon flux from a source by a factor of $e^{-\tau(E,z)}$. 
The most distant of the sources
detected above 100 GeV thus far is PKS 1424+240, with the redshift lower limit
of $z>0.6$ \citep{furniss1424}, which consistently displays a VHE flux of at least
(1.02$\pm$0.08)$\times10^{-7}$ ph cm$^{-2}$~s$^{-1}$ above 120 GeV \citep{archambault}.  
PKS\,1424+240 has been detected at gamma-ray opacities of $\tau>4.5$, 
with the specific value dependent on the EBL model used to estimate the opacity.  When
the unattenuated gamma ray emission from this distant blazar
is reconstructed using current EBL models, the 
spectrum shows a puzzling flattening 
or even upturn at the highest energies \citep{archambault}.

Various scenarios have been explored to account for the spectral shape of high energy emission of 
sources at high opacity values.  These possibilities
include lower EBL densities, as described in \cite{furniss1424}, the
observation of secondary VHE emission resulting from extragalactic 
ultra-high-energy cosmic-rays (UHECR) interactions along the line of sight
(e.g. \citealt{essey2010}), the existence of axion-like particles which couple to the high energy 
gamma rays in the presence of a magnetic field
 \citep{dominguezALPS,hornsmeyer,manuel, rubtsov, chadwick, tavecchio}, and the possibility
that VHE photons avoid pair production with the EBL through gravitational
lensing \citep{barnacka}.    Each of the these scenarios, however, share an important underlying assumption about the density of
the EBL: they assume that, after accounting for the expected cosmological evolution of the photon field, 
at any redshift and wavelength the density of the EBL is homogeneous throughout intergalactic space. 

We present preliminary examination of this assumption in this work.  We perform a simple analysis to investigate a possible relationship 
between extragalactic VHE-like gamma-ray source directions and the density of matter along 
the line of sight.  In our analysis, VHE-like gamma-ray sources are sources that pass criteria on gamma-ray flux as measured by the $Fermi$ Large Area Telescope (LAT; \citealt{Atwood2009}).  See Section 3 for details on the VHE-like gamma-ray source list included in this study.
We find an indication that VHE-like gamma-ray emitting galaxies (more specifically, blazars) might lie preferentially along underdense intergalactic lines of sight.  
We discuss the possibility that VHE-like gamma-ray sources are correlated with directions of 
lower than average EBL density, as this would result in a path of lower gamma-ray opacity for extragalactic gamma rays.  
%Finally,
%we attempt to explain this result in terms of EBL inhomogeneity
%as well as other physical scenarios such as low spatial diffusion of intergalactic pair cascades in the presence of lower magnetic field strength along the blazar lines of sight.

\section{Method}
%Although the EBL is produced by sources at all distances, we investigate the effect 
%on gamma ray attenuation by cosmic voids along the line of sight to distant blazars.
Since the galaxies that are sources of the EBL
preferentially trace high-density regions within the cosmic
web, the local light production
%this is NOT an appropriate reference: \citep{Kaiser1984} 
should be proportional to the local matter density.  
%We extend this to the assumption that, 
Thus, regions with lower matter
density, known as voids, should correlate with regions of lower local light production.   
We take the void sample from the Public Cosmic Void
Catalog\footnote{\url{http://www.cosmicvoids.net}}~\citep{Sutter2012a};
specifically the \emph{2014.11.04} release, as the tracer for regions of
underdensity within the Universe, and investigate the prevalence of voids
along the lines of sight to blazars that lie within the SDSS DR7 survey region. 

The voids are identified
within four volume-limited subsamples of the SDSS DR7 main
sample~\citep{Strauss2002} from redshift $z=0.05$ to $0.2$ and a single
volume-limited subsample from the SDSS LRGs~\citep{Eisenstein2001} from
$z=0.16$  to $z=0.36$. These correspond to the \emph{dim1}, \emph{dim2},
\emph{bright1}, \emph{bright2}, and \emph{lrgdim} samples,  respectively.  For
this analysis we do not take higher-redshift voids from SDSS DR9 because that
survey has a more complicated geometry than the main sample, making the construction
of fairly-distributed random points difficult (see below).  However, for two
particularly distant sources, PKS~1424+240 and PG~1553+113, we utilize this
higher redshift void catalog from SDSS DR9.  Since the DR9-derived void catalog 
used in the analysis of these two sources is not used elsewhere, we do not
include the results for these sources in the population investigations.  When 
looking at different populations of sources, we utilize all voids within
the catalog, including those identified near the survey boundaries, since for
our purposes we only wish to estimate the underdensity fraction along a given
line of sight, and do not need accurate distributions of void shapes. 

Voids in the Public Catalog are identified with the {\tt VIDE}
toolkit~\citep{Sutter2014d}, which uses {\tt ZOBOV}~\citep{Neyrinck2008} to
perform a Voronoi tessellation to estimate the density field from the galaxy
population and a watershed transform to merge regional basins into voids.   The Voronoi tessellation is a segmentation of the volume by a set of polyhedra such that the polyhedra describe the local volumes of each galaxy.  Hence these polyhedra (called Voronoi cells) can be used to estimate the density field of the galaxy population. {\tt
VIDE} ensures that identified voids do not extend outside the survey mask. In
the watershed picture, a void is a non-spherical aggregation of Voronoi
volumes~\citep{Platen2007} and can in principle have any mean or minimum
density. However, they do represent basins in the density
field~\citep{Sutter2013b}, and thus we can use them to estimate the underdense
fraction. While the cores of watershed-based voids have densities below 0.2 $\rho_{mean}$, at the effective radius (see below), the spherically averaged density inside a void is roughly 0.8 $\rho_{mean}$ \citep{Sutter2012a}. 

{\tt VIDE} provides two essential definitions used throughout this work.  The
first is the effective radius of a void, 

\begin{equation}
  R_{\rm eff} \equiv \left( \frac{3}{4 \pi} V \right)^{1/3},
\label{eq:reff}
\end {equation}

\noindent where $V$ is the total volume of all the Voronoi cells that make up
the void.  The second is the \emph{macrocenter}, or volume-weighted  center of
all the Voronoi cells in the void:

\begin{equation}
  {\bf X}_v = \frac{1}{\sum_i V_i} \sum_i {\bf x}_i V_i,
\label{eq:macrocenter}
\end{equation}

\noindent where ${\bf x}_i$ and $V_i$ are the positions and Voronoi volumes of
each galaxy $i$, respectively.

The voids have a typical size (median) of $R_{\rm eff} \approx
20 h^{-1}$Mpc, but range anywhere from $5$ to $60 h^{-1}$Mpc.  For simplicity,
   we approximate the voids as spheres with $R=R_{\rm eff}$ and place them on
the sky according to their macrocenters.

We parameterize the relative underdensity along the Line of Sight (LoS) to an
extragalactic source through the ``voidiness"  ($V_{\rm LoS}$), which is
the fraction of the comoving LoS distance through underdense regions of space
(voids) as compared to the total LoS distance to a source.  More specifically, 

\begin{equation}
V_{\rm LoS}=\frac{\sum D_{\rm void \textit{i}}}{D_{\rm total}}
\label{eq:vlos}
\end{equation}
 
\noindent where $D_{\rm void \textit{i}}$ is the comoving intersectional LoS distance
through a single void $i$ and $D_{\rm total}$ is the total comoving LoS
distance to the source.  A higher $V_{\rm LoS}$ corresponds to a LoS that
intersects a greater fraction of voids, while a low
$V_{\rm LoS}$ corresponds to a lower fraction of underdense intersection.

As a cross-check, we calculate $D_{\rm void}$ in two independent ways.  The first
is a more simplistic method that associates a source to a void if the radial
distance between the source and the void is less than the void radius 
after projecting the source and the void on the sky.
For voids with $z_{void}<z_{source}$, the $D_{\rm void ~\textit{i}}$ is estimated as $(4/3)
R_{\rm eff}$, which is the average length of random
intersections through a spherical region of radius $R_{\rm eff}$ \citep{coleman}. 
We also calculated $D_{\rm void}$ using an alternative approach that
discretized each line connecting the observer to the source into many
($\sim$1000) points. We then obtained $V_{\rm LoS}$ by counting the fraction of
line segments that lay within $R_{\rm eff}$ of each void macrocenter. This more
careful and more expensive calculation resulted in no significant  differences.
Here, we present the detailed results utilizing the former method.

\section{Source Catalogs}
\begin{deluxetable*}{lccc}
\tablewidth{0pc}
\tablecaption{Source catalogs included in this study. See text for a more detailed description of each.}
\tablehead{
 \colhead{Catalog}    &\colhead{Reference}   & \colhead{Number  of  Sources\tablenotemark{*}} & \colhead{Energy Range}  
  %\colhead{Name}  & \colhead{}     & \colhead{in SDSS DR7 Suvey Region}   \\
%  \colhead{}  & \colhead{}     & \colhead{\&}  \\
%  \colhead{}  & \colhead{}     & \colhead{$0.05<z<0.36$}     
}
\startdata
SDSS Quasars&  \cite{SDSS}   & 3901 & optical\\  
2LAC &  \cite{2lac}&  51 &100 MeV$- $100 GeV \\
1FHL &  \cite{1fhl}& 28 &$>10$ GeV\\
VHE-like& \cite{1fhl}  &19  &$>50$ GeV\tablenotemark{**}\\
Randomly Generated & -- &$1\times10^5$ &--
\enddata
\tablenotetext{*}{Sources within SDSS DR7 Survey Region and having a redshift of $0.05<z<0.36$.}
\tablenotetext{**}{Which have gamma-ray properties that pass the following criteria, predetermined within the 1FHL catalog: (1) the flux above 50 GeV being greater
than 1$\times10^{-11}$ ph cm$^{-2}$s$^{-1}$, (2) a spectral index $\Gamma$
above 10 GeV of less than 3 and (3) a measure of the source significance above
30 and 100 GeV (Sig$_{30}= \sqrt{\rm{TS}_{30-100}+\rm{TS}_{100-500}} > 3$).  See text for details.}
\end{deluxetable*}

The analysis is completed for multiple catalogs of both gamma-ray emitting and
non-emitting extragalactic sources.  The catalogs included are summarized in Table~1.  
Our studies are restricted to the catalog sources that 
are within the SDSS DR7 survey region and have a spectroscopically determined
redshift of $0.05<z<0.36$.   The source populations are compared to a random set
of points uniformly distributed throughout the SDSS DR7 survey region.
We also investigate the lines of sight for the sources within the SDSS quasar catalog, which
includes confirmed quasars from the SDSS Seventh data release \citep{SDSS}, as well as 
%The SDSS quasar catalog includes all quasars that have luminosities larger
%than $M_i = -22.0$ and are not characteristically gamma-ray emitterswithin 
the second $Fermi$
catalog of active galactic nuclei (2LAC; \citealt{2lac}).  

The catalog of sources detected by the $Fermi$ LAT above 10 GeV
is referred to as the 1FHL catalog \citep{1fhl}.   This catalog is built using analysis which focuses on energies above 10 GeV.  %The 10 GeV minimum energy
%used for the 1FHL catalog analysis is a good compromise between having an
%adequate number of photons measured by the LAT and being close to the energy
%range where the EBL-VHE photon pair production cross-section is non-zero.
The number of 1FHL sources that have associations with VHE
emitters detected by imaging atmospheric Cherenkov telescopes (IATCs) is 84.  This is a significant portion of the 123 known VHE emitters at the time of 1FHL publication.   
Notably, the majority of the sources with VHE
associations not included in the 1FHL catalog are spatially extended Galactic VHE sources.  There are only five point-like
VHE sources that are not in 1FHL catalog: the starburst
galaxies NGC 253 and M82 and the blazars 1ES~0414+009, 1ES~0229+200,
1ES~1312-423.  None of the aforementioned blazars fall within the SDSS
DR7 survey region and therefore are not included in this study despite their
VHE-detection.   

We explicitly choose not to use the VHE catalog\footnote{\tt http://tevcat.uchicago.edu/} directly in this study.
The ground-based
detections of VHE-emitting sources by IACTs are typically obtained through pointed
observations triggered by elevated emission of a source, and therefore likely result in a high-state biased population.  This bias could be misleading when looking at correlations between $V_{\rm LoS}$ and the gamma-ray flux from a source, as done in Section 4.  For this reason, we also remove 1ES\,0806+524 from this study, the \textit{only} 1FHL source within the SDSS DR7 survey region noted to display significant variability above 10 GeV.

We calculate the $V_{\rm LoS}$ for all 1FHL sources in the SDSS DR7 survey region with $0.05<z<0.36$.  We also pay special attention to
sources that pass a subset of cuts on gamma-ray emission properties to highlight candidate VHE sources that had not yet been
detected by ground-based IACTs.  The cuts were defined within the 1FHL catalog.   The 1FHL sources are flagged as good VHE
candidates based on three properties:  (1) the flux above 50 GeV being greater
than 1$\times10^{-11}$ ph cm$^{-2}$s$^{-1}$, (2) a spectral index $\Gamma_{10}$
above 10 GeV of less than 3 and (3) a measure of the source significance above
30 and 100 GeV (Sig$_{30}= \sqrt{\rm{TS}_{30-100}+\rm{TS}_{100-500}} > 3$),
where TS$_{30-100}$ and TS$_{100-500}$ are the test statistic (TS) values\footnote{See \cite{mattox} for TS definition, a statistic used in \textit{Fermi} LAT source analysis.} from the 30-100 GeV
and 100-500 GeV energy bands.   
%3C~279 was detected at VHE energies during a flare
%that occurred before the launch of the $Fermi$ LAT \citep{3C279}.  Moreover, this
%source is at $z=0.536$, beyond the cosmological distance covered by the void catalog and therefore
%is not included in this study despite the TeV-detection history.

As this VHE candidacy criteria selects all VHE-detected sources at $0.05<z<0.36$, we refer to this population as ``VHE-like" for the remainder of this work.  
%With the inclusivity of extragalactic VHE-detected objects, we believe that
%VHE-like sources are a good tracer for VHE sources.  
A full
list of all 1FHL sources within the SDSS DR7 survey region and that have spectroscopic
redshift $0.05<z<0.36$ is provided in Table~2.  Note that VHE-like source list includes both VHE-detected and VHE-candidate sources, denoted with a Y and C in column 5 of Table~2, respectively.  Additionally, the 1FHL integral gamma-ray flux ($F_{10}$) and index ($\Gamma_{10}$) are also included.

\begin{deluxetable*}{lcccccccc}
\tablecolumns{8}
\tablewidth{0pc}
\tablecaption{1FHL sources $0.05<z<0.36$ within the SDSS DR7 footprint, ordered by increasing $V_{\rm LoS}$. Sources marked with ``Y" are VHE-detected, sources marked with ``C" are VHE-candidates and sources marked with ``N" are not VHE-detected and do not pass the VHE-candidate cuts summarized in Section 3.}
\tablehead{
\colhead{Name} & \colhead{RA}   & \colhead{Dec } & \colhead{Redshift}  & \colhead{VHE} & \colhead{F$_{10} \times10^{-11}$ } & \colhead{$\Gamma_{10}$}  & \colhead{$V_{\rm LoS}$} & \colhead{In Void?}\\
\colhead{} & \colhead{}   & \colhead{} & \colhead{}  & \colhead{Detection?} & \colhead{[ cm$^{-2}$s$^{-1}$]}  & \colhead{} & \colhead{} & \colhead{}
}
\startdata

RX J1136.5+6737 &173.939&67.612&0.13&Y&11.7$\pm$3.3&1.98$\pm$0.34&0.00&No\\
PKS 0829+046 & 128.000 & 4.484& 0.17 & N & 11.9$\pm$ 4.1&2.61$\pm$0.57&0.00&No\\
RX J0847.1+1133 & 131.773&11.544&0.2&Y&9.6$\pm$3.6&1.45$\pm$0.39&0.00&No\\ 
RGB J2313+147&348.514&14.769&0.16&C&10.6$\pm$3.8&1.77$\pm$0.39&0.00&No\\
 H 1426+428&217.158&42.659&0.13&Y&21.2$\pm$4.8&2.01$\pm$0.29&0.00&No\\
PG 1437+398&219.835&39.555&0.35&N&5.9$\pm$2.6&1.79$\pm$0.49&0.02&No\\
PKS 1509+022&228.064&2.074&0.22&N&11.1$\pm$4.1&2.98$\pm$0.70&0.03&No\\
1RXS J115404.9-001008 &178.525&-0.169&0.25&N&11.6$\pm$4.1&2.21$\pm$0.54&0.03&No\\
PMN J1256-1146 & 194.110&-11.77 &0.06&C&11.9$\pm$4.1&2.07$\pm$0.44& 0.05&No\\
1ES 1440+122&220.737&11.982&0.16&Y&7.8$\pm$3.5&1.77$\pm$0.47&0.20&Yes\\
1ES 0806+534\tablenotemark{*} & 122.461 & 52.294 & 0.14& Y & 45.3$\pm$  6.7 & 2.14$\pm$ 0.20 &0.21& No\\
SBS 1200+608&180.807&60.471&0.07&C&5.3$\pm$2.4&1.63$\pm$0.49&0.25&No\\ 
 RBS 0958   &169.305&20.227&0.14&C&23.4$\pm$5.4&1.94$\pm$0.28&0.29&No\\
OJ 287 & 133.712 & 20.079 & 0.31 &N &11.1$\pm$3.9&2.68$\pm$0.59&0.30&Yes\\
MS 1458.8+2249&225.275&22.639&0.23&N&17.2$\pm$4.7&1.89$\pm$0.32&0.30&No\\
ON 246&187.586&25.377&0.14&N&7.3$\pm$3.1&3.29$\pm$0.95&0.41&No\\
RX J1136.8+2551 &174.267&25.893&0.16&C&7.3$\pm$3.1&1.62$\pm$0.44&0.49&Yes\\
MS 1221.8+2452&186.146&24.628&0.22&C&7.7$\pm$3.1&1.26$\pm$0.38&0.49&No\\
BZB J1417+2543&214.659&25.658&0.24&N&8.4$\pm$3.4&2.62$\pm$0.70&0.53&No\\
H 1013+498 &153.773 & 49.427 & 0.21 & Y & 78.7$\pm$8.9&2.28$\pm$0.16&0.54&Yes\\
 RX J1100.3+4019&165.165&40.315&0.23&C&11.2$\pm$3.8&2.23$\pm$0.45&0.60&Yes\\
B2 1229+29&187.970&28.824&0.24&C&28.3$\pm$5.9&2.47$\pm$0.33&0.61&No\\
W Comae&185.402&28.247&0.1&Y&34.1$\pm$6.4&2.18$\pm$0.26&0.62&No\\
GB6 J1053+4930 &163.403 &49.521&0.14 &C & 7.7$\pm$ 2.9&1.49$\pm$0.37&0.62&No\\
 TXS 1055+567&164.666&56.459&0.14&C&44.4$\pm$6.5&2.02$\pm$0.19&0.72&No\\
PG 1218+304&185.337&30.194&0.18&Y&57.2$\pm$8.6&1.90$\pm$0.18&0.78&No\\
B2 1147+24 &177.711&24.320&0.2&N&7.2$\pm$3.0&4.14$\pm$1.28&0.83&Yes\\
 1ES 1215+303  &184.460&30.104&0.13&Y&51.6$\pm$8.2&2.11$\pm$0.21&0.83&Yes\\
\enddata
\tablenotetext{*}{Removed from analysis as the single source in this study showing a Variability Index of 2, as noted in the 1FHL catalog \citep{1fhl}. }
\end{deluxetable*}

\section{Results}
The restrictions on the spatial location and cosmological distance
limits the sources from each catalog included in this study.  The number of sources within these restrictions are summarized in Table~1 for each catalog.
For the 1FHL sources, Table~2 includes the $V_{\rm LoS}$
of each source, as well as information on which sources reside within
a void.  The sources from each catalog are distributed similarly within the 
survey volume.
More specifically, 7 of the 28 1FHL sources  reside within a void (25\% of 1FHL sources).  Five
of the 19 VHE-like sources are within voids (26\%).  These
fractions are consistent with the fraction of SDSS quasars that are found within a void (710 of
3901; 18\%).  Similarly, the randomly distributed points have a 21\% in-void fraction. 

\begin{figure}
\includegraphics[scale=0.45]{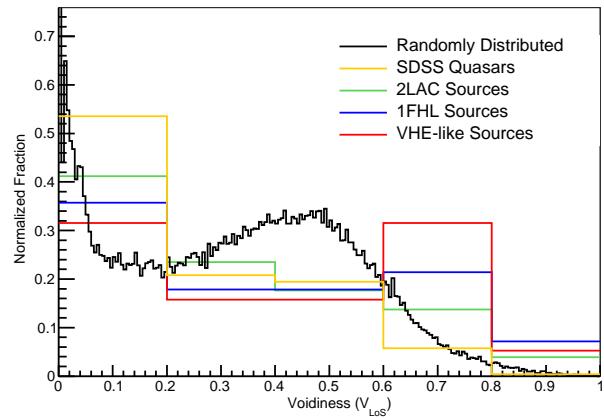}
\caption{Normalized histograms of ``voidiness'' $V_{\rm LoS}$ (Eq.~\ref{eq:vlos}) for the catalog populations (Randomly generated: black, SDSS: yellow, 2LAC: green, 1FHL: blue, VHE-like: red).  See Table 2 for KS test results.}
\end{figure}

Figure~1 shows the normalized histograms of source $V_{\rm LoS}$ for the various
catalogs.  The $V_{\rm LoS}$ of the randomly generated sample of points 
peaks at low $V_{\rm LoS}<0.2$, with the fraction of sources decreasing with greater
$V_{\rm LoS}$.  Less than 1\% of the randomly distributed
sources show $V_{\rm LoS}>0.8$. The slight peak in the distribution of randomly generated sample points
near $V_{\rm LoS} \sim 0.5$ is due to the underlying redshift distribution of the 
voids. 

The SDSS quasars follow a
pattern in the $V_{\rm LoS}$ distribution that is similar to the randomly distributed sample of points. The SDSS quasars, however,
have a higher fraction of sources at low $V_{LoS}<0.2$ (55\% as compared to
42\% for randomly distributed sources).  The number of SDSS quasars with
$V_{\rm LoS} >0.8$ is less than 1\%, again similar to the randomly distrubted sample of points.

\begin{deluxetable*}{c|cccc}
\tablewidth{0pc}
\tablecaption{Cross-correlation of voidiness for different source populations.  KS-test results include P-values as well as the correpsonding significance.}
\tablehead{
\colhead{ } & \colhead{Random}   &\colhead{SDSS}    & \colhead{2LAC}  & \colhead{1FHL} }% \colhead{Candidates}   }
\startdata
%\hline
%               & Random                                   & BAT                                                  & 2LAC                             & 1FHL                & Candidates
Random     &1         &                        &&\\  %            &7.4$\times10^{-4}$ (4.2$\sigma$)     &0.63 (0.7$\sigma$)                        &0.18(1.9$\sigma$)     &6.3$\times10^{-2}$ (2.5$\sigma$) \\
SDSS &  1.6$\times10^{-35}$(12.5$\sigma$)     &          1      &&\\%        &4.5$\times10^{-5}$ (4.1$\sigma$)&0.4 (0.9$\sigma$)&\\
%BAT            &7.3$\times10^{-4}$ (3.4$\sigma$)      &  4.5$\times10^{-5}$ (4.1$\sigma$)  &1                         && \\         %       & 0.56 ( 0.9$\sigma$)                   & 0.30 (1.5$\sigma$)       &0.61 (0.7$\sigma$)     \\
2LAC          &0.63 (0.5$\sigma$)                  &       0.4 (0.9$\sigma$)        & 1        & \\    %                       & 0.90 (0.3$\sigma$)       &0.53 (0.9$\sigma$) \\
1FHL          &0.13 (1.5$\sigma$)                   &    6.2$\times10^{-2}$ (1.7$\sigma$)      &0.83 (0.2$\sigma$)                     &1            \\    %       &0.999(2.3$\sigma$)\\
VHE-like&3.9$\times10^{-2}$ (2.1$\sigma$)    &    1.8$\times10^{-2}$ (2.4$\sigma$)  &0.39 (0.9$\sigma$)                      &  0.999 (0.0$\sigma$)     
\enddata
\end{deluxetable*}

The $V_{\rm LoS}$ distributions of the 2LAC sources is similar to the randomly
generated and SDSS quasars, although not as steeply peaked at low 
$V_{\rm LoS}$.  However, the 1FHL and VHE-like distributions of $V_{\rm LoS}$ are dissimilar 
to the random distribution in
shape: these distributions tend to prefer higher values of 
$V_{\rm LoS}$.  For a more rigorous statistical test of independence 
between each $V_{\rm LoS}$ distribution, we performed a
Kolmogorov-Smirnov (KS) test, with the
P-values and the corresponding significances summarized in Table~3.   The
VHE-like $V_{\rm LoS}$  distribution is marginally different than the 
distribution from
the randomly distributed points: we can reject the null hypothesis at
the 2.1$\sigma$ level that these two samples are drawn from the same 
distribution. A similar significance holds compared to the SDSS quasars
(2.4$\sigma$), while the 2LAC $V_{\rm LoS}$ distribution is 
consistent with both the random and SDSS distributions (at 0.5$\sigma$ and
0.9$\sigma$, respectively).  Less significant is the 1FHL $V_{\rm LoS}$
distribution independence from the random and SDSS distributions at 1.5$\sigma$
and 1.7$\sigma$, respectively. It is not entirely surprising that there is no
significant independence between the 2LAC, 1FHL and VHE-like distributions of
$V_{\rm LoS}$ as there is significant overlap among these source
lists.

%We highlight that the fraction of 1FHL sources to 2LAC sources, catalogs which
%are known to have significant source overlap, is highest at the largest $V_{\rm
%LoS}$ ($>80\%$) and lowest and smallest $V_{\rm LoS}<20\%$ $\sim 45\%$.  More
%specifically, 2LAC sources which are at high $V_{\rm LoS}$ (1ES 1215+303 and B2
%1147+24) have been significantly detected above 10 GeV.  This fraction
%systematically decreases as $V_{LoS}$ decreases, and is summarized in Table~3.
%Unfortunately, with the limited number of sources, this trend is interesting,
%but not statistically significant. 

The results of the KS tests, and more specifically, the independence of the 
$V_{\rm LoS}$ distributions for the VHE-like sources as compared to the randomly generated
and SDSS $V_{\rm LoS}$ distributions, suggest that the distribution of the
VHE-like gamma-ray sources may be related to the intervening matter
density in a different way than the quasar distribution.   We 
investigate whether the level of intervening material might be related to the
detection of gamma-ray emission by looking for a correlation between the
$V_{\rm LoS}$ and the gamma-ray flux above 10 GeV for the VHE-like sources, as shown in Figure~2.  
A constant fit to the $V_{\rm LoS}$ versus F$_{10}$ results 
in a $\chi^2$ of 175 for 17 degrees of freedom, not providing a good representation, especially at the higher values of $V_{\rm LoS}$.

\begin{figure}
\includegraphics[scale=.44]{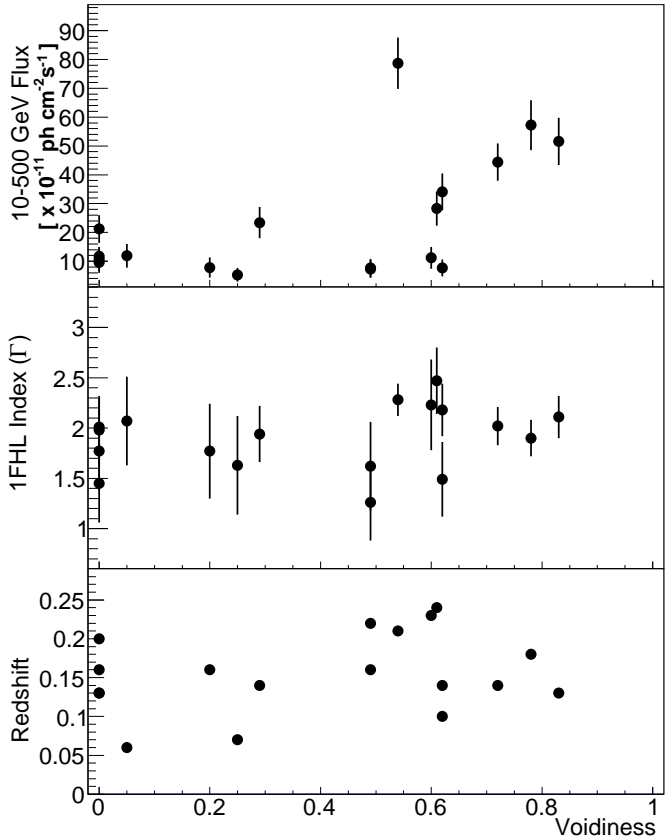}
\caption{$V_{\rm LoS}$ of the VHE-like sources as compared to the source hard gamma-ray flux F$_{10}$ (top panel), the 1FHL index $\Gamma_{10}$ (middle panel), and the source redshift (bottom panel). }
\end{figure}

The Pearson correlation coefficient ($r$) is a measure of the linear dependence between two variables.  This statistical test shows an indication of correlation between
$V_{\rm LoS}$ and the gamma-ray flux of the VHE-like sources above 10 GeV (F$_{10}$), while there is no evidence of a correlation between $V_{\rm LoS}$ and the source index $\Gamma_{10}$ or redshift.
More specifically, for the VHE-like sources $r$ is 0.57 (2.4$\sigma$; 18 sources) for $V_{\rm LoS}$
versus F$_{10}$, 0.25  for $V_{\rm LoS}$ versus $\Gamma_{10}$ and 0.2
for $V_{\rm LoS}$ versus redshift.     For the
1FHL sources, there is also a marginal  correlation between $V_{\rm LoS}$ and the source hard gamma-ray flux with a $r$ value of 0.47 ($\sim2.6\sigma$; 28 sources) when testing $V_{\rm LoS}$
versus F$_{10}$.  The $r$ value is only 0.2 when testing $V_{\rm LoS}$ versus $\Gamma_{10}$ and 0.0
when testing $V_{\rm LoS}$ versus redshift.

%\subsection{Two Exceptionally Distant Steady TeV Blazars}
%PKS 1424 and PG 1553 - what is voidiness, is it consistent with our hypothesis that EBL density is lower in cosmic voids?

\section{Discussion}
We present a preliminary investigation on the assumption that the EBL density is homogeneous.  The indication that the projected location of extragalactic VHE-like sources is related to the intervening matter density aligns with the hypothesis that the EBL density is spatially inhomogeneous.  Only the local contribution to the spatial density of EBL photons is expected to scale with the local number density of galaxies.  This local contribution cannot be more than a small fraction of the total EBL, which mostly arises from distant sources. Each spherical shell around any point contributes approximately equally, since the $r^{-2}$ fall-off of light is compensated by the $r^2$ increase in area; contributions from higher redshifts suffer from redshifting and dilution by the expansion of the universe.  However, since the opacity is an integral along the entire line of sight from a blazar, even a relatively small decrease in the local EBL density could have a cumulative effect sufficient to affect the total gamma-ray opacity significantly.  
%the detection of extragalactic gamma-ray sources probing higher gamma-ray opacities (as estimated by a model-dependent EBL density average) does not challenge the underlying assumptions about cosmological formation history of stars and galaxies, which are a fundamental aspect of EBL model production and evolution estimates.  Moreover, an inhomogeneous EBL density 
This could remove the motivation to exploit scenarios that utilize secondary VHE emission produced along the line of sight or hypothetical axion-like particles to describe the VHE spectral flattening of the intrinsic emission from distant VHE gamma-ray sources such as PKS\,1424+240.

Motivated by the apparent correlation between VHE-like sources and the fraction of void-space along the line of sight, we use another component of the public void catalog that extends to $z=0.65$ (derived from the SDSS DR9 survey region) to investigate the $V_{\rm LoS}$ of the exceptionally distant VHE-detected sources PKS\,1424+240 at $z>0.6035$ and PG\,1553+113 at $z>0.43$.  Both of these sources are within the SDSS DR7 and DR9 survey regions and are noted in the 1FHL catalog as VHE detected sources that display VHE-like gamma-ray flux qualities.  Under the assumption that the sources are located at the redshift lower limits, we find the $V_{\rm LoS}$ for PKS\,1424+240 and PG\,1553+113 are 0.74 and 0.63, respectively.  Due to the use of an additional void catalog in the calculation of the $V_{\rm LoS}$ for these sources, we do not include these sources in the statistical tests performed on the full catalogs, but the relatively high values of $V_{\rm LoS}$ for each are consistent with our findings from the DR7 survey region alone.

We estimate the change in opacity for extragalactic VHE photons traveling along lines of sight that intersect many voids through the investigation of an extreme case.   With the simplifying assumption that a gamma-ray emitting source %at $z\le1$ 
lies along a sightline surrounded by a 50$h^{-1}$ Mpc radius tunnel that is devoid of galaxies, we find that the EBL density within this tunnel is only about 2\% lower than the nominal EBL density, as predicted indepencdently by both the \cite{gilmore2012} and \cite{dominguez} models.  %This slight decrease in density is not compeltely unexpected since the majority of the $z<1$ EBL is produced at $z>1$.  
The effect on the attenuation of gamma rays will be slightly larger than 2\% since the attenuation for gamma rays is derived from the exponent $\tau$.   More specifically, the attenuation decreases by approximately $\tau\times p$, where $p$ represents the percentage decrease in EBL density.  For example, for an opacity of $\tau\sim5$, as probed through the detection of PKS\,1424+240 at the highest gamma-ray energies, the decrease in attenuation scales by roughly 5 $\times$ 2\%=10\%.  This is not sufficient to account for the marginal spectral hardening which appears in the intrinsically emitted VHE spectrum above $\sim300$ GeV.  The red symbols in Figure 3 represent the intrinsic spectrum for the decreased attenuation estimated to result from a 2\% decrease in EBL density.   The results of the calculations suggest that the \textit{anisotropic} location of VHE-like sources with respect to intervening voids cannot be entirely accounted for by a lower EBL photon density inside a void as compared to outside a void.  

\begin{figure}
\includegraphics[scale=.44]{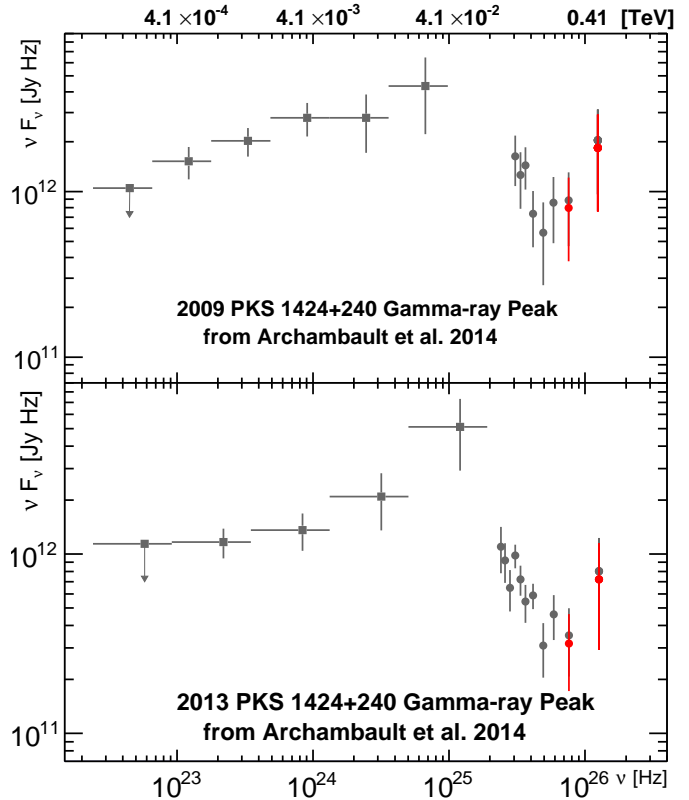}
\caption{The absorption-corrected gamma-ray peak of PKS\,1424+240 from \cite{archambault} as corrected by the \cite{gilmore2012} model (black points) and for the $\tau\times p$ lower absorption resulting from the $p=2\%$ lower EBL density (red points).  Below 300 GeV, the black and red points are indistinguishable and therefore are left as black.}
\end{figure}

An alternative explanation of the distribution of extragalactic VHE-like sources is dependent on the magnitude of the magnetic field within intervening voids.  Proposed astrophysical mechanisms for generating large-scale magnetic fields predict little magnetic flux inside void regions \citep{beck}, while primordial generation mechanisms can create arbitrarily large void magnetic fields, though the latter mechanisms are somewhat constrained by CMB polarization observations (see \citealt{carilli} for a review).  With a low intervening magnetic field within voids, the pair cascades resulting from EBL and VHE photon interaction and/or cosmic ray interaction would remain close to the direct line of sight.  The pairs will often have sufficient energy to upscatter CMB and EBL photons to gamma-ray energies, as described in \cite{neronov}.  Observation of gamma-ray variability above and below the gamma-ray opacity of $\sim$2 can be used to test the feasibility of secondary cascade emission contributing a significant portion of the observed gamma-ray flux.  %If the EBL density is indeed uniform, our results indicate negligible magnetic fields inside voids, which prefer astrophysical generation mechanisms such as dynamos in active galactic nuclei.  One implication of the low magnitude of the magnetic field within intervening voids is that the extragalactic gamma-ray sources with a significant fraction of the line of sight intersecting intervening voids would no longer be feasible targets for indirectly probing the magnitude of the intergalactic magnetic field through observation of gamma-ray point source diffusion (see, e.g., \citealt{dolag}).

In summary, this work shows preliminary but intriguing indications for a non-trivial relationship between the properties of VHE-like sources and the line-of-sight void intersection.  The investigation of the distribution of sources as related to the fraction of intervening underdense regions hints that VHE-like sources are located preferentially along underdense lines of sight.  This same correlation does not appear for randomly distributed points, SDSS quasars or lower-energy gamma-ray AGN.  Moreover, the percentage of sources that fall within voids is consistent across the randomly distributed, SDSS quasar, 2LAC, 1FHL and VHE-like populations.  In other words, all source types are distributed in a similar fashion within the cosmic web, but only VHE-like sources are correlated on the sky with low-density lines of sight.  Expanding this study to include more sources and voids will take years of observational effort but is likely necessary to determine if the marginal correlation found here is a statistical fluctuation.  The results presented here, not accounted for by regions of lower EBL density, suggest that VHE-detection of extragalactic sources may not solely depend on the intrinsic VHE emission from a source, but instead depends on a convolution of the source VHE emission and location within the cosmic web.%characteristics of a gamma-ray sshould be used as motivation for future extragalactic IACT discovery observations to consider not only the multiwavelength characteristics as motivation to select VHE-candidate blazars, but also the location of the source with respect to $V_{\rm LoS}$ values.  

%The fact that simulations show only modest EBL spatial variation, and hence cannot explain the correlations observed in this work, motivate further VHE observations of gamma-ray sources to understand alternative explanations of this marginal .  %and investigation of variability patterns above and below $\tau=2$. 
%Further analysis, such as blah blah tim arlenÕs stuff, blah blah something about more detailed density estimations, and blah blah maybe something else will further elucidate this potential relationship and provide constraints on EBL densities and the origins and nature of large-scale magnetic fields.

\acknowledgments

We thank Rudy Gilmore for discussions.  PMS acknowledges support from NSF Grant NSF AST 09-08693 ARRA. This work made in the ILP LABEX (under reference ANR-10-LABX-63) was supported by French state funds managed by the ANR within the Investissements dÕAvenir programme under reference ANR-11-IDEX-0004-02.  JRP acknowledges support from NSF-AST-1010033.  A.~Dom\'inguez ackowledges the support of the Spanish MICINN's Consolider-Ingenio 2010 Programme under grant MultiDark CSD2009-00064.

\clearpage

%\begin{figure}
%\epsscale{.80}
%\plotone{f1.eps}
%\caption{Derived spectra for 3C138 \citep[see][]{heiles03}. Plots for all sources are available
%in the electronic edition of {\it The Astrophysical Journal}.\label{fig1}}
%\end{figure}

%\begin{deluxetable}{l|ccc}
%\tablewidth{0pc}
%\tablecaption{Fraction of 2LAC sources which are 1FHL sources, broken down by $V_{\rm LoS}$.}
%\tablehead{
% \colhead{$V_{\rm LoS}$}       & \colhead{Number     } & \colhead{Number}  & \colhead{Percentage} \\
%  \colhead{Bin}       & \colhead{of 2LAC     } & \colhead{of 1FHL}  & \colhead{[\%]} }
%\startdata
%0.8-1.0     & 2  &2&100\\  
%0.6-0.8   &  7 &6&85\\  
%0.4-0.6   & 9  &5&55\\  
%0.2-0.4   &12  & 5  &42\\
%0.0-0.2&   21&10&45
%\enddata
%\end{deluxetable}
%
\end{document}